\documentclass[twocolumn,showpacs,preprntnumbers,amssymb]{revtex4}
\usepackage{graphicx}
\usepackage{dcolumn}
\usepackage{bm}
\begin{document}
\title{Encryption with Synchronized Time-Delayed Systems} 
\author{Won-Ho Kye}
\author{Muhan Choi}
\author{Chil-Min Kim}
\affiliation{National Creative Research Initiative Center for Controlling Optical Chaos,
Pai-Chai University, Daejeon 302-735, Korea}
\author{Young-Jai Park}
\affiliation{Department of Physics, Sogang University, Seoul 121-742, Korea}

\begin{abstract}
We propose a new communication scheme that uses time-delayed 
chaotic systems with delay time modulation.
In this method, the transmitter encodes a message as an additional modulation of the delay time
and then the receiver decodes the message by tracking the delay time. 
We demonstrate our communication scheme in a system of coupled logistic maps.
Also we discuss the error of the transferred message due to an external noise 
and present its correction method. 
\end{abstract}

\maketitle

The fact that a chaotic system, which intrinsically 
posses unpredictability and a broad band spectrum, can be synchronized \cite{SyncOrg0,SyncTut, SyncBook}
has led people to assume that a chaotic system would provide better security in communication 
than other cryptographic schemes previously proposed.
Since the first demonstration of its possibility in an electronic circuit \cite{CommCircuit}, 
chaos communication has been extensively investigated and  
correspondingly various methods for masking the message have been developed \cite{Comm}.
It was shown, however, that the message, when masked by a chaotic signal from a low-dimensional 
chaotic system, can be extracted \cite{Extract}.
Further, it was reported that even when the message is masked by a hyperchaotic signal, it can be 
extracted by using nonlinear dynamic forecasting  
as far as the local dynamics does not reflect more 
complicated dynamics significantly \cite{Short}. 

A cure for the weakness of the conventional chaos communication
was to develop methods using synchronization of time-delayed systems \cite{TD_Sync,TD_Comm,TD_Comm1,Hale,TD_Laser,TD_Bio}. 
Despite of the small number of physical degrees of freedom, 
the time-delayed system
has the advantage of possessing the property of high-dimensional hyperchaos which can be easily implemented electronically \cite{TD_Comm}. 
This made the communication using time-delayed systems attract much attention \cite{TD_Comm,TD_Comm1}.
However, it was again reported that  
delay time $\tau$ can be also detected by analyzing the transmitted signal  
and that in such a  case the reconstructed phase space of the 
time-delayed system collapses into low-dimensional manifold \cite{Read_Delay}. 
On this discovery, it was demonstrated that the message masked by 
the signal of time-delayed system can be extracted even 
in the presence of message signal of small amplitude \cite{TD_ATT}.
So far almost all of the methods proposed for chaos communication 
have been broken by their successive counter examples. 
This fact has spurred a debate on the general 
assumptions on a chaotic system.

Meanwhile, a time-delayed system with delay time modulation (DTM) 
in which the delay time is driven by chaotic or stochastic signal 
for the purpose to make the delay time undetectable was introduced \cite{DTM}.
It was analyzed that a phase space reconstruction is
hardly possible in that case \cite{Char_DTM} because DTM significantly increases 
the complexity of an attractor and it renders the delay time undeterminable. 
The report that robust synchronization can be established between two coupled chaotic systems with DTM 
calls attention to a time-delayed system with DTM as an ideal candidate for communication.
%coupled coupled chaotic system in the presence 
%of delay time modulation (DTM), in which the delay time is modulated by 
%another system, can establish the robust synchronization state \cite{SyncTut,SyncBook}.
In the synchronization with DTM, the system
consists of two parties.
One is a transmitter, which is a time-delayed system with delay time modulation:
\begin{eqnarray}
\dot{x}&=&f(x(t), x(t-\tau)), \nonumber \\
\tau &=& g(x(t), t),	
\end{eqnarray}
where $g(x(t), t)$ is the modulation function. 
The other is a receiver which is a time-delayed system driven by 
the signal $x(t-\tau)$ from the transmitter:
\begin{equation}
\dot{x}^\prime=f(x^\prime(t), x(t-\tau)).
\end{equation}  
Even though the modulation works only at the transmitter side, 
it has been found, the receiver is synchronized with the transmitter
by the signal $x(t-\tau)$ \cite{DTM}.

In this Letter, we propose a communication method that 
hides a message into the delay time of time-delayed systems with DTM.  
In our method the transmitter encodes the message as an 
additional modulation of the delay time by which
we overcome the drawbacks in the previous schemes.
The receiver decodes the message 
by identifying the modulated delay time with a delay buffer. 

First, we shall describe the proposed scheme conceptually and demonstrate it in two coupled logistic maps. 
The transmitter is given by:
\begin{eqnarray}
	\dot{x}&=&f(x, x(t-\tilde{\tau})), \nonumber\\
	\tau&=&g(x), \nonumber\\
	\tilde{\tau}&=&\tau+m(t). 
\end{eqnarray} 
Where we call $\tau$ the bare delay time which is modulated by the function $g(x)$ and
$\tilde{\tau}$ the genuine delay time which includes the message $m(t)$. 
Here we start by supposing that the $g(x)$ depends only on the state variable $x$,
and that $g(x)$ is announced in public.
Even if random people know $g(x)$, they can not get bare delay time $\tau$ 
because for that they also need to know the synchronized state variable $x$.
Now we consider the case that the transmitter and receiver are already synchronized after some transient time.
For communication, the transmitter sends $x^* \equiv x(t-\tilde{\tau})$ 
which includes the encoded message.
We emphasize here that our encoding method  
is fundamentally different from the previous ones.
Up to now, in the conventional methods \cite{Comm} the encoding of the message was usually done
by perturbing the real trajectory of the chaotic signal such that $\bar{x}=x+m$.
Accordingly, the message could be extracted by identifying the fluctuations of the reconstructed
trajectory on phase space or return map \cite{Short}. 
Moreover, because the message plays a role of effective noise which degrades 
the quality of synchronization, the system should 
have fast relaxation to the attractor to keep the communication quality,
which potentially increases vulnerability \cite{Comm,Extract}.
On the other hand, in our scheme, the message is encoded as a modulation 
of the delay time such that $\bar{x}=
x(t-(\tau + m))$ which is just the temporal shifting of the original trajectory depending on the message $m$.
Therefore the amplitude of a message need not to be small. 
Furthermore, since the delay time including the message changes the genuine characteristics of the attractor
like the embedding dimension and the complexity through the modulation of the delay time \cite{Char_DTM}, 
the message is constituted into the attractor itself rather than a perturbation.  
%Furthermore, there is no limitation in the power of the message signal.

To decode the message, the receiver should preserve the history of its own 
trajectory for the time interval $ [t-\tau_{m}, t]$ in the delay buffer which is
denoted by the symbol $\{x^\prime [i] \}^t_{t-\tau_m}$, where $\tau_{m}=\max \tau(t)$,  $i\in [0, N]$.
Here $N=[\tau_m/\delta t]$ where $\delta t$ is a sampling step, and $[x]$ is the largest integer less than $x$.
For tracking the delay time, the receiver defines the delay identification measure like this:
\begin{equation}
	M(i, \epsilon) = \epsilon -|x^*-\{x^\prime [i]\}_{t-\tau_m}^t|,
\end{equation}
where $\epsilon$ is the predefined threshold for the identification and $|\cdot|$ 
is the absolute value.
The receiver can find the value of $\tilde{\tau}$
by finding the index $i^*$ which maximizes the delay identification measure 
such that $M(i^*, \epsilon) \geq M(i, \epsilon)$ for $i \in [0, N-1]$.
%$\|x^*-\{x^\prime [i]\}_{t-\tau_m}^t\| \leq\epsilon$ 
%for $i \in [0, N]$, where $\epsilon$ is the decision parameter.
%For simplicity, we shall denote the process such that: 
%$\tilde{\tau}=\|x^*- \{x^\prime [i] \}_{t-\tau_m}^t\| /\epsilon$.
Accordingly, the two delay times, i.e., bare and genuine delay times, can be obtained following the procedure:
\begin{eqnarray}
	\dot{x^\prime}&=&f(x^\prime, x(t-\tilde{\tau})), \nonumber\\
	\tau^\prime&=&g(x^\prime),	\nonumber\\
	\tilde{\tau}^\prime&=&\frac{i^*}{N}\tau_m, 
\end{eqnarray} 
and then the message hidden in the delay time can be decoded at the receiver side such that: 
\begin{equation}
         m^\prime(t)=\tilde{\tau}^\prime-\tau^\prime. 	
\end{equation}
Since we have supposed that, the two systems in our scheme are synchronized, 
the state variables and the delay times are coincided 
with each other, i.e., $x=x^\prime$, $\tau=\tau^\prime$ and $\tilde{\tau}=\tilde{\tau}^\prime$.
Thus the decoded message at the receiver side $m^\prime(t)$ is equal to the original message $m(t)$.

In conventional schemes, since the message is encoded as a deviation
from the synchronized trajectory, the encoded message can be
partly decoded through the synchronized windows in such cases 
as a different system is located on the intermittently synchronized phase. 
However, in our scheme, even if a different system is located on the intermittently synchronized phase,
the message is completely unknown, because the message is not a deviation from the
synchronization manifold. 
%Only the system, which is completely synchronized, can decode the message.
Another strength of this scheme is that the delay time is represented by the integer
even if one use a chaotic flow for transmitter and receiver, because delay time is actually an indicator 
of a position of the previous state.
For that reason, digital data can be directly 
encoded and decoded in this scheme. 

\begin{figure}
\begin{center}
\rotatebox[origin=c]{0}{\includegraphics[width=8.0cm]{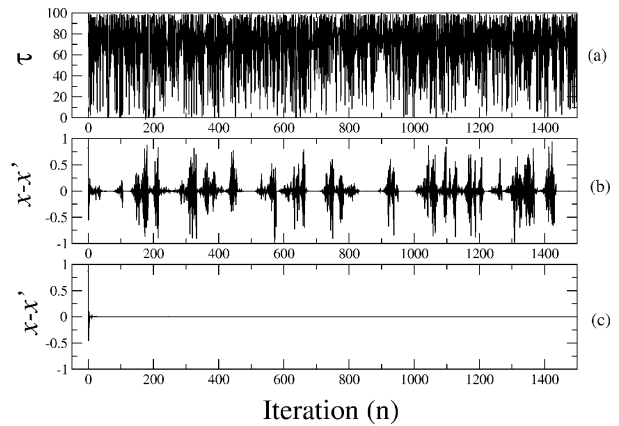}}
\caption{Temporal behaviors of the coupled systems in Eq. (7) and (8), when $\Lambda=100$. 
(a) The modulated delay time; The difference of two state variables: $x-x^\prime$  
(b) below the threshold $\alpha=0.28$ (c) above the threshold $\alpha=0.33$. }
\end{center}
\end{figure}

For demonstration, we consider two logistic maps:
\begin{equation}
\left.
\begin{array}{lll}
x_{n+1}&=& \lambda \bar{x}_n (1-\bar{x}_n),\nonumber\\
\tau &=& [\Lambda x_n], \nonumber\\
\tilde{\tau}&=&\tau + m ~~\mbox{mod}~\tau_m ,  \\
\end{array}
\right \} \mbox{Transmitter}
\end{equation}
where we take $g(x)=[\Lambda x]$ 
and
\begin{equation}
\left.
\begin{array}{lll}
x^\prime_{n+1} &=& \lambda \bar{x}^\prime_n (1-\bar{x}^\prime_n), \nonumber\\
\tau^\prime &=& [\Lambda x^\prime_n], \nonumber\\
\tilde{\tau}^\prime&=&i^*, \nonumber \\
m^\prime&=&\tilde{\tau}^\prime-\tau^\prime ~~\mbox{mod}~\tau_m, 	
\end{array}
\right \} \mbox{Receiver}
\end{equation}
where  $\bar{x}_n=(1-\alpha) x_n + \alpha x_{n-\tilde{\tau}}$ and  
$\bar{x}^\prime_n=(1-\alpha) x^\prime_n + \alpha x_{n-\tilde{\tau}}$ and we take $\lambda=4.0$.
Here $i\in [0, \tau_m-1]$ and $i^*$ is the identified index which maximizes the delay identification measure (Eq. (4)).
% which satisfies 
%$\epsilon-|x^*-\{x^\prime[i^*]\}^{n}_{n-\tau_m}| \geq \epsilon-|x^*-\{x^\prime[i]\}^{n}_{n-\tau_m}|$ 
Figure 1 shows the temporal behaviors of the two coupled systems with a null message, i.e., $m=0$.
While the difference $x-x^\prime$ between the two state variables 
shows the intermittent chaotic bursting below the synchronization threshold (Fig. 1(b)),  
it converges fast to the synchronization manifold $x-x^\prime=0$ 
above the threshold (Fig. 1(c)). 

\begin{figure}
\begin{center}
\rotatebox[origin=c]{0}{\includegraphics[width=8.0cm]{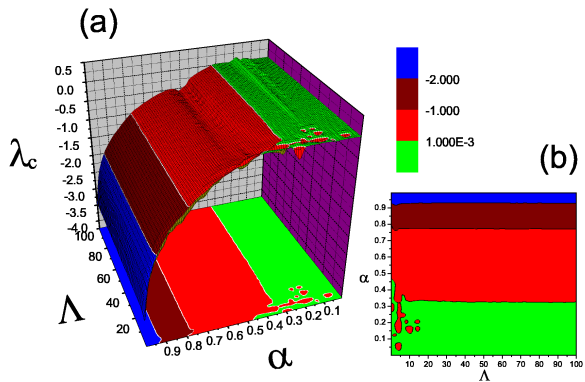}}
\caption{(a) The conditional Lyapunov exponent $\lambda_c$ as a function of $\Lambda$ and $\alpha$.
	(b) The contour plot in the $(\Lambda, \alpha)$ space. 
	Here the colors indicate the different values of $\lambda_c$. }
\end{center}
\end{figure}

To analyze the critical behaviors of the two coupled chaotic systems we find the 
difference motion such that:
$\Delta X_{n+1}=\lambda (1-\alpha)(1-(\bar{x}_n + \bar{x}_n^\prime))\Delta X_n$
where $\Delta X_n=x_n-x^\prime_n$.
The conditional Lyapunov $\lambda_c$, 
which determines the synchronization threshold,  
can be found by following the standard procedure:
$\lambda_c=\lim_{n\rightarrow \infty} \frac{1}{n} \log |\Delta X_n/\Delta X_0|$.
\begin{figure}
\begin{center}
\rotatebox[origin=c]{0}{\includegraphics[width=8.0cm]{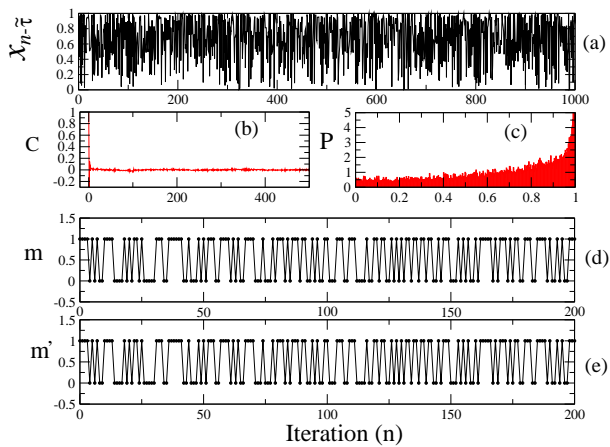}}
\caption{ (a) The transmitted signal including the message $m$ at $\alpha=0.7$, $\Lambda=100$, and $\epsilon=10^{-8}$; 
	(b) The autocorrelation of the transmitted signal, which shows $\delta$ function shape; 
	(c) The probability distribution function of the transmitted signal $x_{n-\tilde{\tau}}$;
	(d) The original message at transmitter side; (e) The decoded message at receiver side.}
\end{center}
\end{figure}
Figure 2 shows conditional Lyapunov exponent as a function of coupling strength $\alpha$ and
modulation amplitude $\Lambda$. One sees that the conditional Lyapunov exponent becomes
negative above $\alpha =0.32$ for all $\Lambda \in [3, 100]$ 
and it means two coupled systems are synchronized in that regime.
On the other hand, if the modulation amplitude is the relatively small, i.e., $\Lambda < 3$, 
synchronization is established in the stronger coupling, i.e., $\alpha >0.48$ 
and some synchronization islands appear in the regime, $\Lambda \in [3, 18]$ and $\alpha \leq 0.32$
(see the contour plot in Fig. 2 (b)). 

\begin{figure}
\begin{center}
\rotatebox[origin=c]{0}{\includegraphics[width=8.0cm]{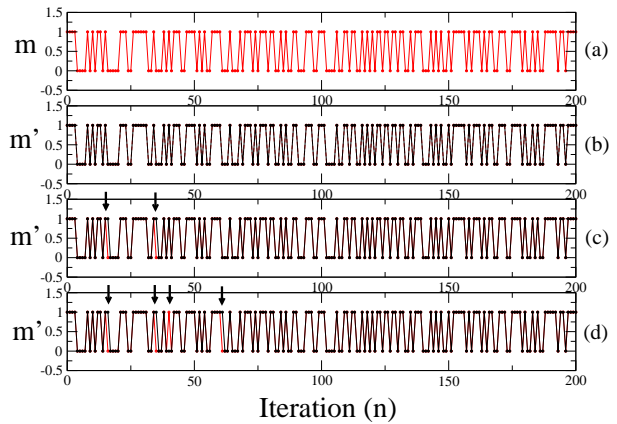}}
\caption{The decoded messages for different values of the threshold $\epsilon$ in the presence of noise of $\max(\xi_n) = 10^{-6}$.
	(a) The original message; (b) The decoded message with $\epsilon=2.5\times 10^{-5}$. 
	(c) With $\epsilon= 2.0 \times 10^{-5}$; (d) With $\epsilon = 1.5\times 10^{-5}$.
	 The arrows in (c) and (d) indicate the bit-flipped events in the decoded message}
\end{center}
\end{figure}
Figure 3 shows the temporal behavior and statistical properties 
of the transmitted signal with an example of the message transfer.
The transmitted signal $x_{n-\tilde{\tau}}$ is shown in Fig. 3 (a) and 
the autocorrelation function of the
signal is presented in Fig. 3 (b).
Figure 3 (c) shows the probability distribution function $P$ of the transmitted signal $x_{n-\tilde{\tau}}$, 
which is a normalized histogram of the projected trajectory onto the $x_{n-\tilde{\tau}}$ axis 
in $n$ versus $x_{n-\tilde{\tau}}$ plot (Fig. 3 (a)).
%To determine $P(x_{n-\tau})$, we simply project the values taken by $x_{n-\tau}$ versus $n$ back onto the $x_{n-\tau}$ axis
%and form a normalized histogram of the values.
Since the autocorrelation function is $\delta$-correlated, 
one can see that the system looks random to an eavesdropper.  
Fig. 3 (d) is the original message encoded at the transmitter
and (e) is the message decoded at the receiver. 
Consideration of the effects of noise in the transmission channel is essential 
in regard to real application and implementation.
The noise in the transmission channel $\xi_n$ induces the distortion at receiver side 
such that:
\begin{equation}
\hat{x}^*=x^*+\xi_n, ~~\hat{x}^\prime_{n} = x^\prime_{n} +O(\alpha\xi_n).
\end{equation}
The distortion is propagated into the delay identification measure 
such that $\hat{M}(i, \epsilon)=\epsilon-|\hat{x}^*-\{\hat{x}^\prime[i]\}^{n}_{n-\tau_m}|$.
So there exists a possibility that the identified index  $i^*$ can be determined 
incorrectly due to the external noise.
%The noise and decision compete with each other in the equation: 
%$\tilde{\tau}^\prime=\|x_{n-\tilde{\tau}}- \{x\}^n_{n-\tau_m} +\xi_n\| /\epsilon$.
Figure 4 shows how the external noise has an effect on the message transfer in our scheme.
The original message is presented in Fig. 4 (a) and 
the decoded messages with different values of the threshold $\epsilon$ for delay time identification 
are presented in Fig. 4 (b)-(d).
One can see that the transferred bits are intermittently flipped as shown in Fig. 4 (c) and (d) (see the arrows).   
%In real implementation, one  need to adjust the decision parameter $\epsilon$ so as to eliminate the  bit flipping
%in the transfered information.
%In case of Fig. 3(b) shows the fact that the message can be completely decoded 
%even in the presence of external noise by
%choosing the decision parameter $\epsilon$, appropriately.    
\begin{figure}
\begin{center}
\rotatebox[origin=c]{0}{\includegraphics[width=8.0cm]{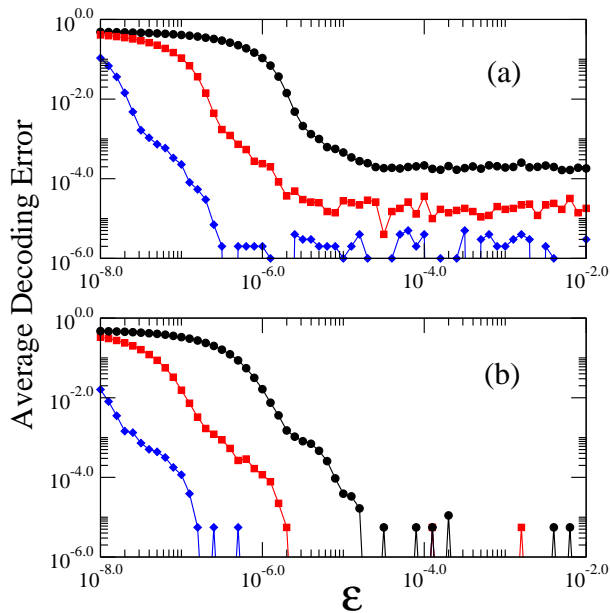}}
\caption{The average decoding errors for different noise amplitudes as a function of the threshold $\epsilon$, 
	where the points below $10^{-6.0}$ are zero. Each datum point is the result of  $10^6$ bit transfer.
	Circles: $\max (\xi) = 10^{-6}$; Squares: $\max(\xi)=10^{-7}$; Diamonds:  $\max(\xi)=10^{-8}$. 
	(a) Without the error correction; (b) With the error correction.
	See Ref. \cite{Err_Corr} for details of the error correction method.}  
\end{center}
\end{figure}
To clarify the role of the noise $\xi_n$ in our scheme and the relationship with 
the threshold $\epsilon$, we evaluate the average decoding error 
which is defined by the number of flipped bits divided by the total number of transferred bits.
Figure 5 (a) shows the decoding error as a function 
of the threshold $\epsilon$ in the presence of external noise.
The decoding error is decreased as the threshold is increased.
Figure 5 (b) shows the data on the application of the error correction method \cite{Err_Corr}. 
With the proper threshold, the receiver can decode 
the message completely even in the presence of external noise.   
 
It is worth discussing the possibility of eavesdropping for the proposed scheme.
The eavesdropper can accumulate the transmitted signal $x(t-\tilde{\tau})$, which 
is the temporally shuffled signal of the synchronized state variable $x(t)$. 
Accordingly the eavesdropper can not reconstruct the phase space correctly \cite{Char_DTM}, because the
information of the phase space is preserved only when the temporal ordering of the signal is kept.
Even if one succeeds in reconstructing the phase space, the message is not 
separable as mentioned above. 
For the eavesdropper to identify the genuine delay time $\tilde{\tau}$, 
the construction of the delay buffer is the most essential procedure.
The eavesdropper, however, always gets 
the fake delay buffer $\{\tilde{x}[i]\}^{t}_{t-\tau_m}$ 
in which the temporal order is mixed.
Accordingly the eavesdropper read the 
incorrect genuine delay time $\tilde{\tau}$ from it.
Furthermore, since the synchronized state variable $x$ is never exposed outside 
the transmitter or the receiver in our scheme, the eavesdropper can not find  
the bare delay time $\tau$ also, which is the function of synchronized state variable $x$.
 
In conclusion, we have proposed a new communication scheme which 
enables one to encode and decode the desired message using the
time-delayed system in the presence of delay time modulation.
Our scheme is fundamentally different from the conventional ones in that, in our scheme,
since the message is encoded as a modulation of the delay time, it does not
degrade the quality of the synchronization;
in the conventional ones the message is encoded 
as a perturbation of the real trajectory of a chaotic system and 
so the quality of synchronization is degraded, 
which eventually increases the potential vulnerability of whole communication systems.
Also the message is decoded just by comparing the 
transmitted signal with the history of the trajectory stored in the delay buffer. 
We have also shown that the scheme works even in 
the presence of  external noise just with  a simple algorithm for the error correction.
We expect our scheme can be used to implement the real communication system 
with better performance and enhanced security. 

\acknowledgments
This work is supported by Creative Research Initiatives 
of the Korean Ministry of Science and Technology.


\begin{thebibliography}{150}
%\bibitem{Huy} Ch. Huygens, {\it Horologium Oscillatiorium} Apud F. Muguet, Paris, France, 1673 English
%		translation: {\it The Pendulum Clock} Iowa State University Press, Ames, 1986.
\bibitem{SyncOrg0} H. Fujisaka and T. Yamada, Prog. Theor. Phys. {\bf 69}, 32 (1983);
		   V.S. Afraimovich, N.N. Verichev, and M.I. Rabinovich, Radiophys. 
		   Quantum Electron. {\bf 29}, 747 (1986);
		    L.M. Pecora and T.L. Carroll, Phys. Rev. Lett. {\bf 64}, 821 (1990).
\bibitem{SyncTut} S. Boccaletti, J. Kurth, G. Osipov, D.L. Valladares and C. Zhou, Physics Reports {\bf 366}, 1 (2002). 
\bibitem{SyncBook} A. Pikovsky, M. Rosenblum, and J. Kurths, {\it Synchronization A universal concept in nonlinear
			science},  CAMBRIDGE UNIVERSITY PRESS, 2001. 
\bibitem{CommCircuit} K.M Cuomo and A.V. Oppenheim, Phys. Rev. Lett. {\bf 71}, 65 (1993).
\bibitem{Comm} L. Kocarev and U. Parlitz, Phys. Rev. Lett. {\bf 74}, 5028 (1995); 
	J. H. Peng, E. J. Ding, M. Ding, and W. Yang, Phys. Rev. Lett. 76, 904 (1996);
	J. H. Xiao, G. Hu, and Z. Qu, Phys. Rev. Lett. {\bf 77}, 2818 (1996);
	S. Boccaletti, A. Farini, and F. T. Arecchi, Phys. Rev. E {\bf 55}, 4979 (1997);
	K. Murali and M. Lakshmanan, Phys. Rev. E {\bf 56}, 251-255 (1997);
	Z. Liu, S. Chen, and B. Hu, Phys. Rev. E {\bf 59}, 2817 (1999);
	S. Sundar and A. A. Minai, Phys. Rev. Lett. {\bf 85}, 5456 (2000);
	C.-M. Kim, S. Rim, and W.-H. Kye, Phys. Rev. Lett. {\bf 88}, 014103 (2002);
	S. Wang, J. Kuang, J. Li, Y. Luo, H. Lu, and G. Hu, Phys. Rev. E {\bf 66}, 065202 (2002);
	R. Mislovaty, E. Klein, I. Kanter, and W. Kinzel, Phys. Rev. Lett. {\bf 91}, 118701 (2003). 
\bibitem{Extract} G. P\'erez and H.A. Cerdeira, Phys. Rev. Lett. {\bf 74}, 1970 (1995).
\bibitem{Short} K.M. Short and A.T. Parker, Phys. Rev. E {\bf 58}, 1159 (1998);
	    D.-U. Hwang, C.-M. Kim, and Y.-J. Park, J. Korean Phys. Soc. {\bf 42}, 8 (2003).
\bibitem{TD_Sync} K. Pyragas, Phys. Rev. E {\bf 58}, 3067 (1998);
		 R. He and P.G. Vaidya, Phys. Rev. E {\bf 59}, 4048 (1999).
\bibitem{TD_Comm} B. Mensour and A. Longtin, Phys. Lett. A {\bf 244}, 59 (1998).
\bibitem{TD_Comm1} 
		 J.-P. Goedgebuer, L. Larger, and Henri Porte, Phys. Rev. Lett. {\bf 80}, 2249 (1998);
		 L. Yaowen, G. Gguangming, Z. Hong, and W. Yinghai, Phys. Rev. E {\bf 62}, 7898 (2000);
	         V.S. Udaltsov, J.-P. Goedgebuer, L. Larger, and W.T. Rhodes, Phys. Rev. Lett. {\bf 86}, 1892 (2001).
\bibitem{Hale} J.K. Hale, {\it Theory of Functional Differential Equations}, Springer-Verlag, Berlin, (1997) and references therein.
\bibitem{TD_Laser} T. Heil, I. Fischer, W. Els\"asser, J. Mulet, and C. R. Mirasso, Phys. Rev. Lett. {\bf 86}, 2001 (795).
\bibitem{TD_Bio}J. Fort and V. M\'endez, Phys. Rev. Lett. {\bf 89}, 178101 (2002).
\bibitem{Read_Delay} R. Hegger, M.J. B\"unner, H. Kantz, and A. Giaquinta, Phys. Rev. Lett. {\bf 81}, 558 (1998).
\bibitem{TD_ATT} C. Zhou and C.-H. Lai, Phys. Rev. E {\bf 60}, 320 (1999);
		 B.P. Bezruchko, A.S. Karavaev, V.I. Ponomarenko, and M.D. Prokhorov,
		 Phys. Rev. E {\bf 64}, 056216 (2001);
		V.I. Ponomarenko and M.D. Prokhorov, Phys. Rev. E {\bf 66}, 026215 (2002).
%\bibitem{Death} D.V. Ramana Reddy, A. Sen, and G.L. Johnston, Phys. Rev. Lett. {\bf 80}, 5109 (1998).
\bibitem{DTM}   W.-H. Kye, M. Choi, M.-W. Kim, S.-Y. Lee, S. Rim, C.-M. Kim, and Y.-J. Park, 
		Phys. Lett. A {\bf 322}, 338 (2004).
\bibitem{Char_DTM} W.-H. Kye, M. Choi, S. Rim, M.S. Kurdoglyan, C.-M. Kim, and Y.-J. Park,
	Phys. Rev. E {\bf 69}, 055202(R) (2004).
\bibitem{Err_Corr} As a simplest implementation of error correction, 
		we have introduced a redundance of the transferred data.
		At the transmitter, a bit is represented by five duplicated bits, e.g., 
		"1" is represented by "11111".
		At the receiver, the bit is read as the value 
		that is duplicated more than two times.  
\end{thebibliography}
\end{document}